\documentclass{article}
\usepackage{dcase2020}
\usepackage{amsmath,graphicx,url,times,booktabs, tabularx, xcolor, hyperref}
\usepackage{graphicx}% http://ctan.org/pkg/graphicx

\newcommand{\website}[1]{\textcolor{blue}{#1}}
\title{SONYC-UST-V2: An Urban Sound Tagging Dataset \\with Spatiotemporal Context}

\name{Mark Cartwright$^{1}$\sthanks{We would like to thank all the Zooniverse volunteers who continue to contribute to our project. This work is supported by National Science Foundation awards 1544753 and 1633259.},
       Jason Cramer$^{1}$,
       Ana Elisa Mendez Mendez$^{1}$,
       Yu Wang$^{1}$}
 \secondlinename{
 	   Ho-Hsiang Wu$^{1}$,
       Vincent Lostanlen$^{1,2}$,
       Magdalena Fuentes$^{1}$,
       Graham Dove$^{1}$
       }
 \thirdlinename{
       Charlie Mydlarz$^{1}$,
       Justin Salamon$^{3}$,
       Oded Nov$^{1}$,
       and Juan Pablo Bello$^{1}$
       }
 \address{$^1$ New York University, New York, NY, USA\\ 
         $^2$ Cornell Lab of Ornithology, Ithaca, NY, USA\\
         $^3$ Adobe Research, San Francisco, CA, USA\\
  }
\begin{document}

\ninept
\maketitle

\begin{sloppy}

\begin{abstract}

\end{abstract}

We present SONYC-UST-V2, a dataset for urban sound tagging with spatiotemporal information.
This dataset is aimed for the development and evaluation of machine listening systems for real-world urban noise monitoring.
While datasets of urban recordings are available, this dataset provides the opportunity to investigate how spatiotemporal metadata can aid in the prediction of urban sound tags.
SONYC-UST-V2 consists of 18510 audio recordings from the ``Sounds of New York City'' (SONYC) acoustic sensor network,  including the timestamp of audio acquisition and location of the sensor.
The dataset contains annotations by volunteers from the Zooniverse citizen science platform, as well as a two-stage verification with our team.
In this article, we describe our data collection procedure and propose evaluation metrics for multilabel classification of urban sound tags.
We report the results of a simple baseline model that exploits spatiotemporal information.


\section{Introduction}
\label{sec:intro}

Often in machine listening research, researchers work with datasets scraped from the internet, disconnected from real applications, and devoid of relevant metadata such as when and where the data were recorded. 
However, this is not the case in many real-world sensing applications. 
In many scenarios, we \textit{do} know when and where the data were recorded, and this spatiotemporal context (STC) metadata may inform us as to what objects or events we may expect to occur in a recording. 
Computer vision researchers have already shown that STC is helpful in detecting objects such as animals in camera trap images and vehicles in traffic camera images \cite{beery2020context}. 
We believe STC may also aid in sound event detection tasks such as urban sound tagging, e.g Figure~\ref{fig:task_definition}, by informing us as to what sound events we may expect to hear in sound recordings.
For example, in New York City you are more likely to hear an ice cream truck by the park at 3pm on a Saturday in July than you are by a busy street at rush hour on a Tuesday in January; however, you are more likely to hear honking, engines, and sirens on that Tuesday. 
But, knowledge of a thunderstorm that Saturday afternoon in July would reduce your expectation to hear an ice cream truck and could also help you disambiguate between the noise of heavy rain and that of a large walk-behind saw. 
However, few works have exploited this information for urban sound tagging \cite{cartwright2019tricycle} or even sound tagging in general. We hypothesize that one of the main reasons for this is the lack of available data with audio and temporal and spatial metadata. 

\begin{figure}
  \centering
  \centerline{\includegraphics[width=0.9\linewidth]{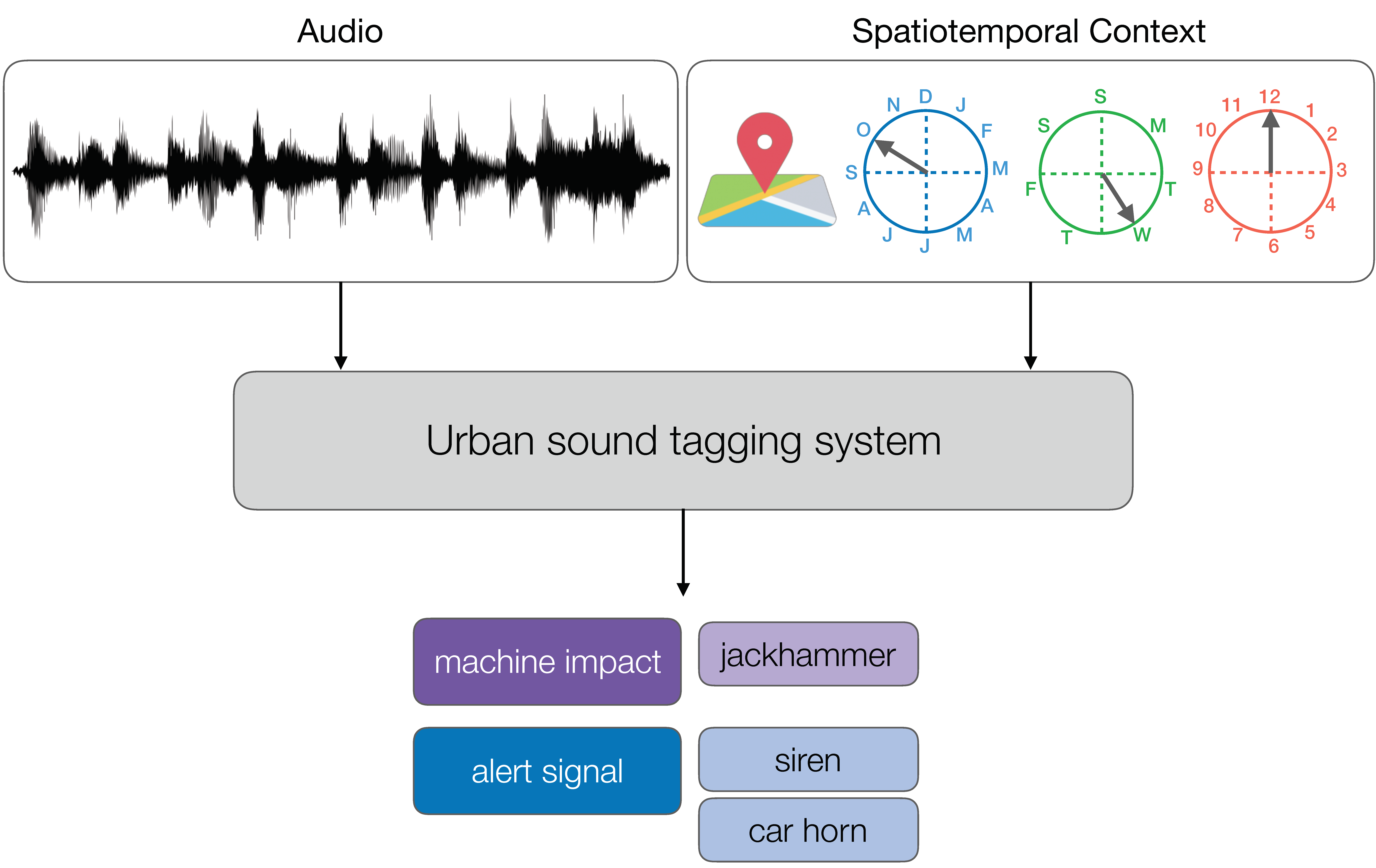}}
  \caption{Overview of a system that exploits spatiotemporal information for urban sound tagging.}
  \label{fig:task_definition}
\end{figure}

In this article, we introduce SONYC-UST-V2, a dataset for urban sound tagging with spatiotemporal information,\footnote{Download the data at \href{https://doi.org/10.5281/zenodo.3966543 }{https://doi.org/10.5281/zenodo.3966543.}} which contains 18510 annotated 10 s recordings from the SONYC acoustic sensor network and which served as the dataset for the DCASE 2020 Urban Sound Tagging with Spatiotemporal Challenge\footnote{\href{http://dcase.community/challenge2020/task-urban-sound-tagging-with-spatiotemporal-context}{http://dcase.community/challenge2020/task-urban-sound-tagging-with-spatiotemporal-context.}}. 
Each recording has been annotated on a set of 23 ``tags'', which was developed in coordination with the New York City Department of Environmental Protection (DEP) and represents many of the frequent causes of noise complaints in New York City. 
In addition to the recording, we provide identifiers for the New York City block (location) where the recording was taken as well as when the recording was taken, quantized to the hour. 
This information alone can be used to help a tagging model learn the ``rhythm'' of the city, but it can also be used query and join external datasets that can provide additional contextual information, e.g. weather, traffic, holidays, land use, city permits, and social data---all of which are available through rich, public datasets.
We hope this data and task can provide a test bed for investigating these ideas for machine listening. 

%In the remainder of the paper, we review the precursor to this dataset (SONYC-UST-V1) and then describe the process of collecting SONYC-UST-V2, the metrics to evaluate models against the ground-truth, and a baseline system trained on both audio and spatiotemporal context.

\website{%This dataset provides resources to investigate how spatiotemporal metadata can aid in the prediction of urban sound tags for 10s recordings from an urban acoustic sensor network for which we know when and where the recordings were taken. This task is motivated by the real-world problem of building machine listening tools to aid in the monitoring, analysis, and mitigation of urban noise pollution.
}

\section{Previous work}
\label{sec:previous_work}

SONYC Urban Sound Tagging (SONYC-UST, referred to from here on as SONYC-UST-V1) is a dataset for the development and evaluation of machine listening systems for real-world urban noise monitoring \cite{cartwright2019sonyc}. It was used for the  Urban Sound Tagging challenge in DCASE 2019, and consists of 3068 audio recordings from the SONYC acoustic sensor network \cite{bello2019sonyc}. This acoustic network consists of more than 50 acoustic sensors deployed around New York City and has recorded 150M+ 10-second audio clips since its launch in 2016. 
The sensors are located in the Manhattan, Brooklyn, and Queens boroughs of New York, with the highest concentration around New York University's Manhattan campus (see Figure \ref{fig:map}). To maintain the privacy of bystanders' conversations and prevent the recording of intelligible conversation, the network's sensors are positioned for far-field recording, 15--25 feet above the ground, and record audio clips at random intervals. 
%SONYC-UST addresses the limitations of previously available datasets for urban sound tagging by providing recordings from realistic urban noise conditions across a variety of times and locations, closely matching the label set to the needs of noise enforcement agencies, and providing annotations in a multi-class fashion. 

The SONYC-UST-V1 dataset contains annotated training, validation, and test splits (2351 / 443 / 274 recordings respectively). These splits were selected so recordings from the same sensors would not appear in both the training and validation sets, and such that the distributions of labels were similar for both the training and validation sets. Finally, the test set is not disjoint in terms of sensors, but rather it is disjoint in time---all recordings in the test set are posterior to those in the training and validation sets. 

The recordings were annotated by citizen volunteers via the Zooniverse citizen science platform \cite{simpson2014zooniverse, zooniverse} and were followed by a two-step verification by our team in the case of the validation and test splits. In Zooniverse, volunteers weakly tagged the presence of 23 fine-grained classes that were chosen in consultation with the New York DEP.
These 23 fine-grained classes are then grouped into eight coarse-grained classes with more general concepts: e.g., the coarse \textit{alert signals} category contains four fine-level categories: \textit{reverse beeper}, \textit{car alarm}, \textit{car horn}, \textit{siren}. Recordings that are most similar to a small set of exemplary clips from YouTube for each sound class in our taxonomy were selected for annotation. We refer the interested reader to \cite{cartwright2019sonyc} for further details about the class taxonomy and the similarity measure used for data selection.
 
\section{Data collection}
\label{sec:data_collection}
 \begin{figure}
  \centering
  \centerline{\includegraphics[width=0.8\columnwidth]{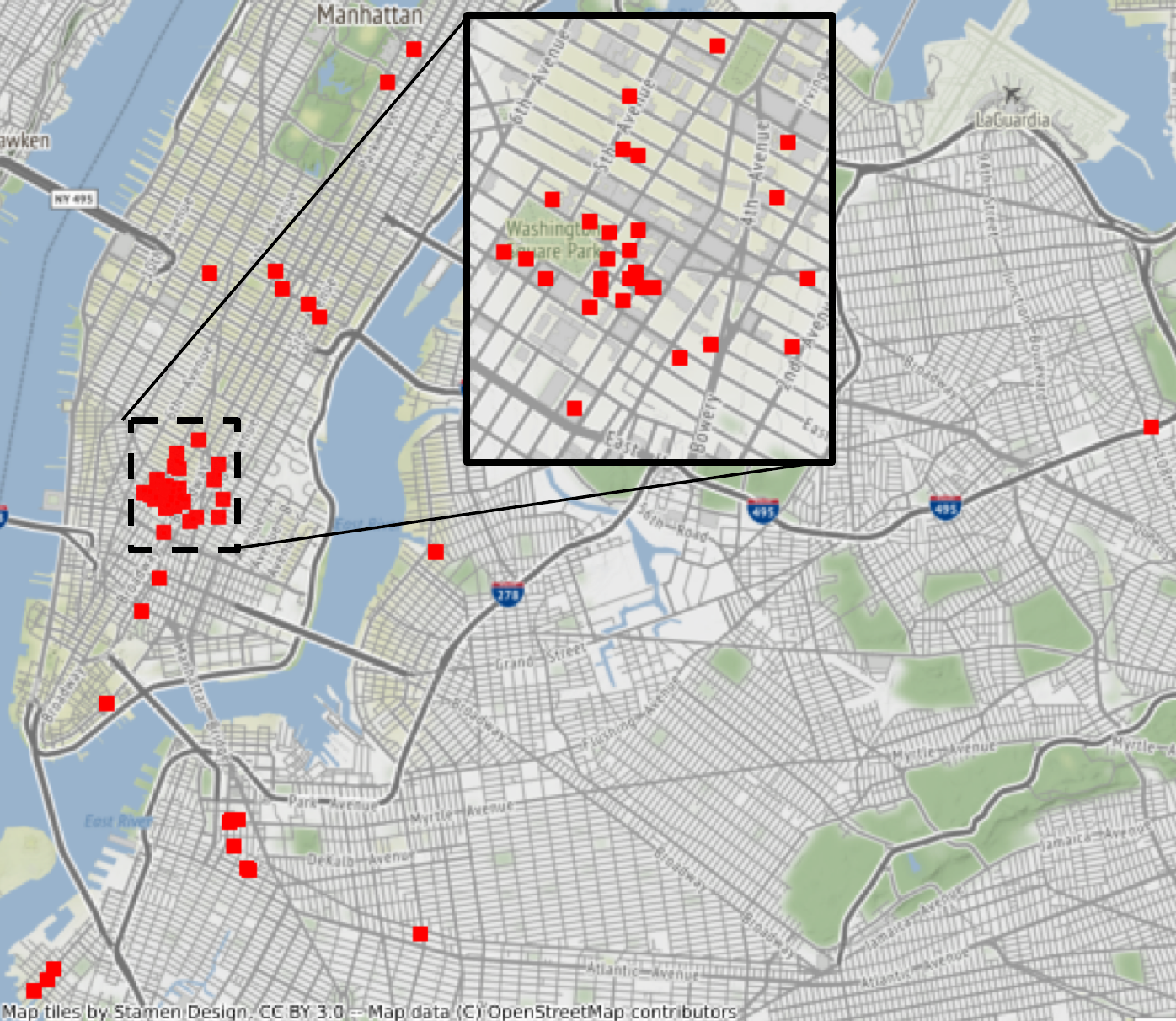}}
  \caption{
  SONYC-UST-V2 sensor locations, many of which are in in Manhattan's Greenwich Village neighborhood (see inset).}
  \label{fig:map}
\end{figure}

Since the release of SONYC-UST-V1, we have continued collecting audio recordings from our acoustic sensor network and Zooniverse volunteers have continued to annotate these recordings. 
SONYC-UST-V2 includes a total of 18510 annotated recordings from 56 sensors, a small sample of the 150M+ recordings that the SONYC acoustic sensor network has collected.
The method for selecting which recordings to annotate has evolved over time. 
Initially, we sampled recordings as we did for V1, i.e., recordings that were most similar to a small set of exemplary clips from YouTube for each sound class in our taxonomy \cite{cartwright2019sonyc}. 
Subsequently, we sampled recordings using a batch-based active learning procedure in which a multi-label classifier was trained with all available annotations at that time. 
The model then predicted the class presence for unlabeled recordings, and recordings with class probabilities above a low fixed threshold were then clustered with minibatch $k$-means \cite{sculley2010web}. 
For each class, recordings were evenly sampled from each cluster to obtain a diverse sample, with more recordings sampled for classes with low representation in the dataset. 
Batch sizes typically varied between 1--2k recordings. 
We sampled the test set with yet another sampling procedure. 
For this set, a random sample of 10k recordings was selected from the set of unlabeled SONYC recordings. 
This was reduced to a diverse subset of 1k recordings selected with a determinantal point process (DPP) using the DPPy package \cite{GPBV19} and OpenL3 embeddings \cite{cramer2019openl3} as the representation. 
This set was reduced further to adhere to our privacy criteria outlined in Section~\ref{sec:STC}.

Each recording in SONYC-UST-V2 has been annotated by three different Zooniverse volunteers in the same manner as SONYC-UST-V1, i.e., on both the presence and proximity of the 23 fine-level and 8 coarse-level urban sound tags from the SONYC-UST Taxonomy \cite{cartwright2019sonyc}.

As in SONYC-UST-V1, a subset of the recordings have annotations verified by the SONYC team in a two-step verification process. 
To create verified labels, we first distributed recordings based on coarse-level sound category to members of the SONYC research team for labeling. 
To determine whether a recording belonged to a specific category for the validation process, we selected those that had been annotated by at least one Zooniverse volunteer. 
Two members of the SONYC team then labeled each category independently. 
Once each member had finished labeling their assigned categories, the two annotators for each class discussed and resolved label disagreements that occurred during the independent annotation process. 
Lastly, a single SONYC team member listened to all of the recordings to ensure consistency across coarse-level categories and to catch any classes overlooked by the crowdsourced annotators.
1380 of the recordings have verified annotations---716 recordings from the SONYC-UST-V1 test and validation sets and 664 new recordings which comprise the SONYC-UST-V2 test set.

In SONYC-UST-V2 we continue our practice of defining training and validation sets that are disjoint by sensor and a test set that is temporally displaced to test generalization in a typical urban noise monitoring scenario. While the dataset contains recordings from 2016--2019, only the test set contains recordings from the latter two thirds of 2019. To capitalize on the effort put into the verified subsets in SONYC-UST-V1, we build upon the existing training and validation sensor split, growing each, while keeping the V1 split still intact. However, the SONYC-UST-V1 test set was not limited to the validation sensor split nor were subsequent crowdsourced annotations limited to recordings in the training sensor split. Thus, we now have verified annotations for recordings in the training sensor split and crowdsourced-only annotations for recordings in the validation sensor split, see Figure~\ref{fig:data_splits}. All of this data has been included for completeness. However, when training the baseline model (see Section~\ref{sec:baseline}), we limit the training set to only the crowdsourced annotations in the training sensor split, and the validation set to only the verified annotations in the validation sensor split. See Figure~\ref{fig:coarse_class_distribution} for the coarse-level class distribution of these recording splits.

Annotating urban sound recordings is a particularly difficult task. Sound events may be very distant with low signal-to-noise ratios, yet still audible. In addition, without visual verification, many sound events can be difficult to disambiguate. To capture this uncertainty, annotators are allowed to provide ``incomplete'' annotations, providing only the coarse-level class when they are unsure of the fine-level class (e.g. ``Other/unknown engine''). Due to this difficult task, the inter-annotator agreement of the crowdsourced annotations as measured by Krippendorff's $\alpha$ \cite{krippendorff2018content} is rather low (0.36). Thus, SONYC-UST-V2 includes all of the individual crowdsourced and verified annotations, and we encourage users of the dataset to explore annotation aggregation strategies that model and incorporate annotator reliability. Since that is out of scope of this article, we use a simple approach of minority vote for our baseline model and analysis, i.e., a class is marked as present in the aggregate if at least one annotator marks it present. In previous work with Zooniverse annotators \cite{cartwright2019crowdsourcing}, we have found this strategy increases recall without significantly decreasing precision. In Table~\ref{tab:results}, we evaluate Zooniverse annotations aggregated with minority vote against the verified annotations in the test set using the metrics outlined in Section~\ref{sec:metric}. These results are likely representative of good model performance when only a simplistic annotation aggregation method is used.

\begin{figure}
  \centering
  \centerline{\includegraphics[width=1\columnwidth]{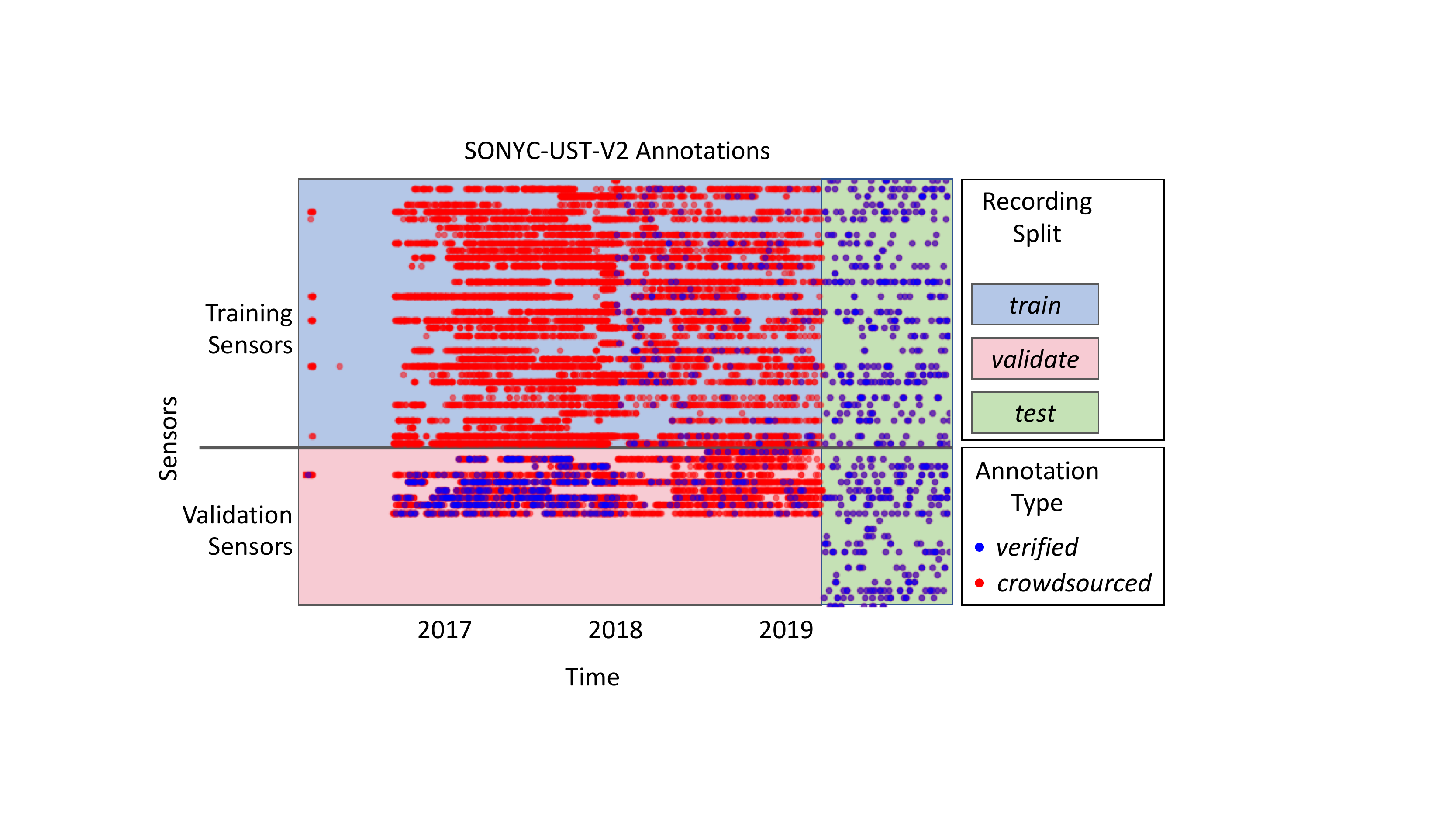}}
  \caption{Dataset splits. 
The sensors in the test set overlap with both the training and validation sets. 
The test data is temporally dislocated from training and validation to test generalizability in time.}
  \label{fig:data_splits}
\end{figure}

\begin{figure}
  \centering
  \centerline{\includegraphics[width=1\columnwidth]{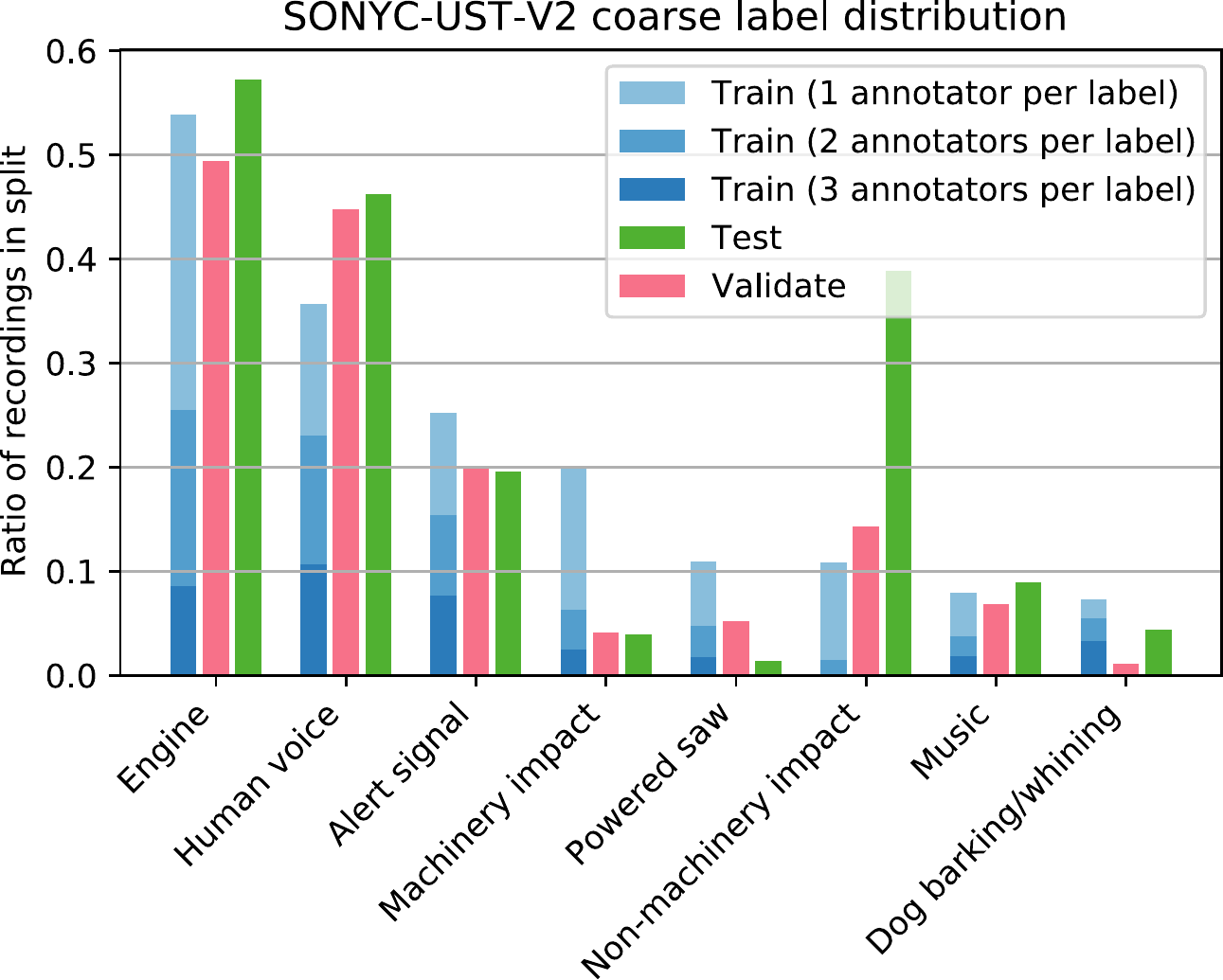}}
  \caption{SONYC-UST tag distribution normalized for each recording split, in decreasing order of frequency in the training split. The shades of blue indicate how many annotators tagged the class in a training set recording, i.e., darker shades of blue indicate higher annotator agreement.}
  \label{fig:coarse_class_distribution}
\end{figure}

\section{Spatiotemporal Context (STC) Information}
\label{sec:STC}

 \begin{figure}
  \centering
  \centerline{\includegraphics[width=\columnwidth]{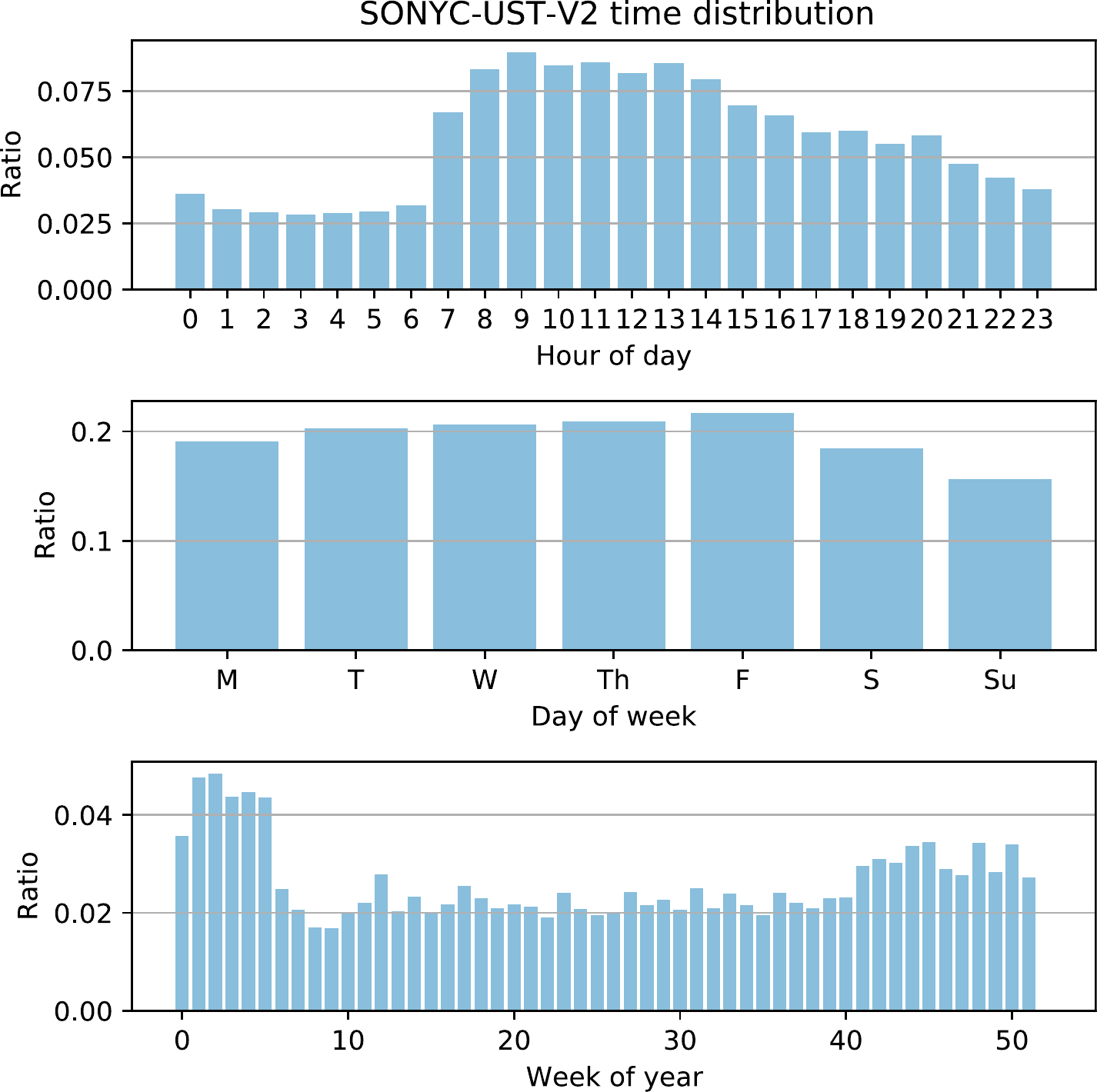}}
  \caption{
  Distribution of dataset recordings per hour of the day, day of the week, and week of the year.}
  \label{fig:temporal_distribution}
 \end{figure}

The unique characteristic of this dataset is the inclusion of spatiotemporal context information, which informs where and when each example was recorded.  To maintain privacy, we quantized the spatial information to the level of a city block, and we quantized the temporal information to the level of an hour. We also limited the occurrence of recordings with positive human voice annotations to one per hour per sensor. For the spatial information, we have provided borough and block identifiers, as used in NYC's parcel number system known as Borough, Block, Lot (BBL) \cite{bbl}. This is a common identifier used in NYC datasets, making it easy to relate the sensor data to other city data such as PLUTO \cite{pluto} and more generally NYC Open Data \cite{nycopendata}, which contain information regarding land use, construction, transportation, noise complaints, and more. For ease of use with other datasets, we've also included the latitude and longitude coordinates of the center of the block. Figures~\ref{fig:temporal_distribution} and \ref{fig:map} are distributions of the recordings in time and space.

\section{Evaluation metrics}
\label{sec:metric}
SONYC-UST-V2 includes labels at two hierarchical levels, coarse and fine (cf.~\cite{cartwright2019sonyc} for details about the taxonomy), and models are evaluated independently against the labels at each level. Since some of the fine-level classes can be hard to label, even for human experts, a fraction of the samples in SONYC-UST-V2 only have coarse labels for some sound events. For example, a distant engine sound may be too ambiguous to label as a \emph{small engine}, a \emph{medium engine} or a \emph{large enging} (i.e., fine labels), but can still tagged with the coarse label \emph{engine of uncertain size}. For such cases, we use a tag coarsening procedure that leverages the hierarchical relationship between the fine and coarse labels in our taxonomy to obtain performance estimates for fine labels in the face of annotator uncertainty (cf.~\cite{cartwright2019sonyc} for further details about this procedure).

For each of the two levels, we compute three metrics: macro-averaged AUPRC, micro-averaged AUPRC, and label-weighted label-ranking average precision (LWLRAP) \cite{fonseca2019audio}. We use the first as the primary performance metric, and the second as a secondary metric to gain further insight into the performance of each system. Macro-averaged AUPRC provides a measure of performance across all classes independently of the number of samples per class, while micro-averaged AUPRC is sensitive to class imbalance. 

Finally, LWLRAP measures the average precision of retrieving a ranked list of relevant labels for each test clip. It is a generalization of the mean reciprocal rank measure for evaluating multi-label classification, which gives equal weight to each label in the test set (as opposed to each test clip). The metric has been widely adopted in the DCASE community over the past year.

% A fraction of the sound tags in SONYC-UST-V2 is impossible to identify precisely, even for human experts.
% This is the case, in particular, when the acoustic source is far away from the sensor.
% Thus, we accommodate label uncertainty by allowing the annotator to supply coarse-level tags, such as \emph{engine of uncertain size} or \emph{other/unknown alert signal}.
% Unlike fine-level tags such as \emph{small engine} or \emph{car horn}, coarse-level tags provide an incomplete ground truth.
% Such incompleteness poses a methodological problem for evaluating multilabel classifiers.

% To address this problem, we propose to evaluate the prediction at the fine level only when possible, and fall back to the coarse level if necessary.
% This operation of tag coarsening allows to evaluate any prediction against the ground truth, resulting in a number of true positives (TP), false positives (FP), and false negatives (FN) in each coarse category.
% We refer to \cite{cartwright2019sonyc} for a mathematical definition of tag coarsening in SONYC-UST, both v1 and v2.

% In each coarse category, integer counts for TP, FP, and FN can be combined into well-known information retrieval metrics: precision, recall, and area under the precision recall curve (AUPRC).
% Furthermore, they can be micro-averaged across coarse categories to yield an overall AUPRC.

\section{Baseline system}
\label{sec:baseline}

\begin{table}
\footnotesize
\begin{center}
\begin{tabular}{l | cc | cc | cc}
\hline
Estimator: & \multicolumn{2}{c |}{Annotators}                          & \multicolumn{2}{c |}{Model w/ STC}                            &
\multicolumn{2}{c}{Model w/o STC}\\
Level: & F             & C            & F           & C          & F             & C          \\
 \hline\hline
\textbf{Overall} & & & & & & \\
\hline
Macro-AUPRC &	0.56 &	0.69 &	0.44 &	0.49 &	0.43 &	0.49 \\	
Micro-AUPRC &	0.60 &	0.75 &	0.62 &	0.71 &	0.62 &	0.71 \\	
LWLRAP &	0.62 &	0.78 &	0.72 &	0.83 &	0.73 &	0.83 \\
\hline
\textbf{AUPRC} & & & & & & \\
\hline
Engine &	0.57 &	0.82 &	0.57 &	0.84 &	0.59 &	0.84 \\
Mach. imp. &	0.35 &	0.48 &	0.19 &	0.32 &	0.18 &	0.30 \\
Non-mach. imp. &	0.60 &	0.60 &	0.58 &	0.60 &	0.59 &	0.61 \\
Powered saw&	0.14 &	0.37 &	0.16 &	0.11 &	0.12 &	0.12 \\
Alert signal &	0.74 &	0.82 &	0.45 &	0.40 &	0.44 &	0.39 \\
Music &	0.53 &	0.75 &	0.41 &	0.52 &	0.41 &	0.54 \\	
Human voice &	0.78 &	0.91 &	0.88 &	0.92 &	0.88 &	0.93 \\	
Dog&	0.79 &	0.79 &	0.26 &	0.22 &	0.24 &	0.23 \\
\hline
\end{tabular}
\end{center}
\caption{The performance of the Zooniverse annotations (using minority vote aggregation) and the baseline classifier with and without STC as compared the the ground-truth annotations for the test split on the coarse (C) and fine (F) levels.}
\label{tab:results}
\end{table}

For the baseline model \footnote{\href{https://github.com/sonyc-project/dcase2020task5-uststc-baseline}{https://github.com/sonyc-project/dcase2020task5-uststc-baseline}}, we use a multi-label multi-layer perceptron model, using a single hidden layer of size 128 (with ReLU non-linearities), and using AutoPool \cite{mcfee2018autopool} to aggregate frame level predictions. The model takes in as input \emph{audio content}, \emph{spatial context}, and \emph{temporal context}.

Audio content is given as OpenL3 \cite{cramer2019openl3} embeddings (with \texttt{content\_type="env"}, \texttt{input\_repr="mel256"}, and \texttt{embedding\_size=512}), using a window size and hop size of 1.0 second (with centered windows), giving us 11 512-dimensional embeddings for each clip in our dataset. Spatial context is given as latitude and longitude values, giving us two values for each clip in our dataset. Temporal context is given as hour of the day, day of the week, and week of the year, each encoded as a one hot vector, giving us 83 values for each clip in our dataset. We z-score normalize the embeddings, latitude, and longitude values, and concatenate all of the inputs (at each time step), resulting in an input size of 597.

We use the weak tags for each audio clip as the targets for each clip. For the training data (which has no verified target), we count a positive for a tag if at least one annotator has labeled the audio clip with that tag (i.e., minority vote). Note that while some of the audio clips in the training set have verified annotations, we only use the crowdsourced annotations. For audio clips in the validation set, we only use annotations that have been manually verified.

We train the model using stochastic gradient descent to minimize the binary cross-entropy loss, using $L^2$ regularization (weight decay) with a factor of $10^{-5}$. For training models to predict tags at the fine level, we modify the loss such that if ``unknown/other'' is annotated for a particular coarse tag, the loss for the fine tags corresponding to this coarse tag are masked out. We train for up to 100 epochs, using early stopping with a patience of 20 epochs using loss on the validation set. We train one model to predict fine-level tags, with coarse-level tag predictions obtained by taking the maximum probability over fine-tags predictions within a coarse category. We train another model only to predict coarse-level tags.

Table~\ref{tab:results} presents the results of the baseline model trained with and without spatiotemporal context. The baseline model's performance is quite low and does not seem to benefit from the inclusion of STC. However, its inclusion of STC and its aggregation of annotations are both rather naive. We hope this simply provides a starting point for researchers to explore more sophisticated approaches that better leverage the unique aspects of this data and incorporate additional contextual data to aid in generalizability. 

\section{Conclusions}
\label{sec:conclusions}
SONYC-UST-V2 is a multi-label dataset for urban sound tagging with spatiotemporal context information. It consists of 18510 audio examples recorded in New York City between 2016 and 2019 with weak (i.e., tag) annotations on urban sound classes, as well as metadata on where and when each audio example was recorded. We believe STC is a rich source of information for sound tagging that has yet to be adequately explored and could potentially aid models in the challenging task of tagging real-world urban sound recordings. This dataset is the first of its kind that we are aware of and will provide researchers with material for exploring the incorporation of spatiotemporal context (STC) information into sound tagging.

% -------------------------------------------------------------------------
% Either list references using the bibliography style file IEEEtran.bst
\bibliographystyle{IEEEtran}
\bibliography{refs}

\begin{thebibliography}{10}
\providecommand{\url}[1]{#1}
\def\UrlFont{\rmfamily}
\providecommand{\newblock}{\relax}
\providecommand{\bibinfo}[2]{#2}
\providecommand\BIBentrySTDinterwordspacing{\spaceskip=0pt\relax}
\providecommand\BIBentryALTinterwordstretchfactor{4}
\providecommand\BIBentryALTinterwordspacing{\spaceskip=\fontdimen2\font plus
\BIBentryALTinterwordstretchfactor\fontdimen3\font minus
  \fontdimen4\font\relax}
\providecommand\BIBforeignlanguage[2]{{%
\expandafter\ifx\csname l@#1\endcsname\relax
\typeout{** WARNING: IEEEtran.bst: No hyphenation pattern has been}%
\typeout{** loaded for the language `#1'. Using the pattern for}%
\typeout{** the default language instead.}%
\else
\language=\csname l@#1\endcsname
\fi
#2}}

\bibitem{beery2020context}
S.~Beery, G.~Wu, V.~Rathod, R.~Votel, and J.~Huang, ``Context r-cnn: Long term
  temporal context for per-camera object detection,'' in \emph{Proceedings of
  the IEEE/CVF Conference on Computer Vision and Pattern Recognition}, 2020,
  pp. 13\,075--13\,085.

\bibitem{cartwright2019tricycle}
M.~Cartwright, J.~Cramer, J.~Salamon, and J.~P. Bello, ``Tricycle: Audio
  representation learning from sensor network data using self-supervision,'' in
  \emph{Proceedings of the 2019 IEEE Workshop on Applications of Signal
  Processing to Audio and Acoustics (WASPAA)}.\hskip 1em plus 0.5em minus
  0.4em\relax IEEE, 2019, pp. 278--282.

\bibitem{cartwright2019sonyc}
M.~Cartwright, A.~E.~M. Mendez, J.~Cramer, V.~Lostanlen, G.~Dove, H.-H. Wu,
  J.~Salamon, O.~Nov, and J.~Bello, ``Sonyc urban sound tagging (sonyc-ust): a
  multilabel dataset from an urban acoustic sensor network,'' in
  \emph{Proceedings of the 2019 Detection and Classification of Acoustic Scenes
  and Events Workshop (DCASE)}, Oct. 2019, pp. 35--39.

\bibitem{bello2019sonyc}
J.~P. Bello, C.~Silva, O.~Nov, R.~L. DuBois, A.~Arora, J.~Salamon, C.~Mydlarz,
  and H.~Doraiswamy, ``Sonyc: A system for monitoring, analyzing, and
  mitigating urban noise pollution,'' \emph{Communications of the ACM},
  vol.~62, no.~2, pp. 68--77, 2019.

\bibitem{simpson2014zooniverse}
R.~Simpson, K.~R. Page, and D.~De~Roure, ``Zooniverse: Observing the world's
  largest citizen science platform,'' in \emph{Proceedings of the 23rd
  International Conference on World Wide Web}, ser. WWW '14 Companion.\hskip
  1em plus 0.5em minus 0.4em\relax New York, NY, USA: ACM, 2014, pp.
  1049--1054.

\bibitem{zooniverse}
\BIBentryALTinterwordspacing
``{Zooniverse}.'' [Online]. Available: \url{www.zooniverse.org}
\BIBentrySTDinterwordspacing

\bibitem{sculley2010web}
D.~Sculley, ``Web-scale k-means clustering,'' in \emph{Proceedings of the 19th
  international conference on World wide web}, 2010, pp. 1177--1178.

\bibitem{GPBV19}
\BIBentryALTinterwordspacing
G.~Gautier, G.~Polito, R.~Bardenet, and M.~Valko, ``{DPPy: DPP Sampling with
  Python},'' \emph{Journal of Machine Learning Research - Machine Learning Open
  Source Software (JMLR-MLOSS)}, 2019, code at
  http://github.com/guilgautier/DPPy/ Documentation at
  http://dppy.readthedocs.io/. [Online]. Available:
  \url{http://jmlr.org/papers/v20/19-179.html}
\BIBentrySTDinterwordspacing

\bibitem{cramer2019openl3}
J.~{Cramer}, H.~{Wu}, J.~{Salamon}, and J.~P. {Bello}, ``Look, listen, and
  learn more: Design choices for deep audio embeddings,'' in \emph{Proceedings
  of the 2019 IEEE International Conference on Acoustics, Speech and Signal
  Processing (ICASSP)}, 2019, pp. 3852--3856.

\bibitem{krippendorff2018content}
K.~Krippendorff, \emph{Content analysis: An introduction to its
  methodology}.\hskip 1em plus 0.5em minus 0.4em\relax Sage publications, 2018.

\bibitem{cartwright2019crowdsourcing}
M.~Cartwright, G.~Dove, A.~E. M{\'e}ndez~M{\'e}ndez, J.~P. Bello, and O.~Nov,
  ``Crowdsourcing multi-label audio annotation tasks with citizen scientists,''
  in \emph{Proceedings of the 2019 ACM CHI Conference on Human Factors in
  Computing Systems}.\hskip 1em plus 0.5em minus 0.4em\relax ACM, 2019, p. 292.

\bibitem{bbl}
\BIBentryALTinterwordspacing
``Borough, block, lot lookup.'' [Online]. Available:
  \url{https://portal.311.nyc.gov/article/?kanumber=KA-01247}
\BIBentrySTDinterwordspacing

\bibitem{pluto}
\BIBentryALTinterwordspacing
``{Property Land Use Tax lot Output}.'' [Online]. Available:
  \url{https://www1.nyc.gov/site/planning/data-maps/open-data.page}
\BIBentrySTDinterwordspacing

\bibitem{nycopendata}
\BIBentryALTinterwordspacing
``{NYC Open Data}.'' [Online]. Available:
  \url{https://opendata.cityofnewyork.us/}
\BIBentrySTDinterwordspacing

\bibitem{fonseca2019audio}
E.~Fonseca, M.~Plakal, F.~Font, D.~P. Ellis, and X.~Serra, ``Audio tagging with
  noisy labels and minimal supervision,'' in \emph{Proceedings of the Detection
  and Classification of Acoustic Scenes and Events 2019 Workshop (DCASE2019)},
  New York University, NY, USA, October 2019, pp. 69--73.

\bibitem{mcfee2018autopool}
\BIBentryALTinterwordspacing
B.~McFee, J.~Salamon, and J.~P. Bello, ``Adaptive pooling operators for weakly
  labeled sound event detection,'' \emph{IEEE/ACM Trans. Audio, Speech and
  Lang. Proc.}, vol.~26, no.~11, p. 2180–2193, Nov. 2018. [Online].
  Available: \url{https://doi.org/10.1109/TASLP.2018.2858559}
\BIBentrySTDinterwordspacing

\end{thebibliography}

%
% or list them by yourself
% \begin{thebibliography}{9}
% 
% \bibitem{dcase2016web}
%   \url{http://www.cs.tut.fi/sgn/arg/dcase2016/}.
%
% \bibitem{IEEEPDFSpec}
%   {PDF} specification for {IEEE} {X}plore$^{\textregistered}$,
%   \url{http://www.ieee.org/portal/cms_docs/pubs/confstandards/pdfs/IEEE-PDF-SpecV401.pdf}.
%
% \bibitem{PDFOpenSourceTools}
%   Creating high resolution {PDF} files for book production with 
%   open source tools, 
%   \url{http://www.grassbook.org/neteler/highres_pdf.html}.
%
% \bibitem{eWilliams1999}
% E. Williams, \emph{Fourier Acoustics: Sound Radiation and Nearfield Acoustic
%   Holography}. London, UK: Academic Press, 1999.
% 
% \bibitem{ieeecopyright}
%   \url{http://www.ieee.org/web/publications/rights/copyrightmain.html}.
%
% \bibitem{cJones2003}
% C. Jones, A. Smith, and E. Roberts, ``A sample paper in conference
%   proceedings,'' in \emph{Proc. IEEE ICASSP}, vol. II, 2003, pp. 803--806.
% 
% \bibitem{aSmith2000}
% A. Smith, C. Jones, and E. Roberts, ``A sample paper in journals,'' 
%   \emph{IEEE Trans. Signal Process.}, vol. 62, pp. 291--294, Jan. 2000.
% 
% \end{thebibliography}

\end{sloppy}
\end{document}